\documentclass[aps,preprintnumbers,superscriptaddress,showpacs]{revtex4}
\usepackage{epsfig}
\usepackage{psfrag}
\usepackage{amsfonts}
\usepackage{graphicx}
\usepackage{dcolumn}
\usepackage{bm}

\begin{document}

\title{Heavy quark potential with hyperscaling violation}

\author{Zi-qiang Zhang}
\email{zhangzq@cug.edu.cn} \affiliation{School of mathematics and
physics, China University of Geosciences(Wuhan), Wuhan 430074,
China}

\author{Chong Ma}
\email{machong@cug.edu.cn} \affiliation{School of mathematics and
physics, China University of Geosciences(Wuhan), Wuhan 430074,
China}

\author{De-fu Hou}
\email{houdf@mail.ccnu.edu.cn} \affiliation{Institute of Particle
Physics and Key Laboratory of Quark and Lepton Physics (MOE),
Central China Normal University, Wuhan 430079, China}

\author{Gang Chen}
\email{chengang1@cug.edu.cn} \affiliation{School of mathematics
and physics, China University of Geosciences(Wuhan), Wuhan 430074,
China}

\begin{abstract}
In this paper, we investigate the behavior of the heavy quark
potential in the backgrounds with hyperscaling violation. The
metrics are covariant under a generalized Lifshitz scaling
symmetry with the dynamical Lifshitz parameter $z$ and
hyperscaling violation exponent $\theta$. We calculate the
potential for a certain range of $z$ and $\theta$ and discuss how
it changes in the presence of the two parameters. Moreover, we add
a constant electric field to the backgrounds and study its effects
on the potential. It is shown that the heavy quark potential
depends on the non-relativistic parameters. Also, the presence of
the constant electric field tends to increase the potential.
\end{abstract}


\maketitle
\section{Introduction}

AdS/CFT \cite{Maldacena:1997re,Gubser:1998bc,MadalcenaReview},
which relates a d-dimensional quantum field theory with its dual
gravitational theory, living in(d+1) dimensions, has yielded many
important insights into the dynamics of strongly-coupled gauge
theories. For reviews, see Refs
\cite{CP,CP1,MA,SS,EC,JS,KB,MC,TM,RG,MA1,UG} and references
therein.

Due to the broad application of this characteristic, many authors
have considered the generalizations of the metrics that dual to
field theories. One of such generalizations is to use metric with
hyperscaling violation. Usually, the metric is considered to be an
extension of the Lifshitz metric and has a generic Lorentz
violating form \cite{TA,KBA,HS,KN,FD}. As we know, Lorentz
symmetry represents a foundation of both general relativity and
the standard model, so one may expect new physics from Lorentz
invariance violation. For that reason, the metrics with
hyperscaling violation have been used to describe the string
theory \cite{BG,EP,MA2,JSA,EK}, holographic superconductors
\cite{EBR,SJS,YYB,QYP} as well as QCD \cite{JSA1,JSA2,KBI}.

The heavy quark potential of QCD is an important quantity that can
probe the confinement mechanism in the hadronic phase and the
meson melting in the plasma phase. In addition, it has been
measured in great detail in lattice simulations. The heavy quark
potential for $\mathcal{N}=4$ SYM theory was first obtained by
Maldacena in his seminal work \cite{Maldacena:1998im}.
Interestingly, it is shown that for the $AdS_5$ space the energy
shows a purely Coulombian behavior which agrees with a conformal
gauge theory. This proposal has attracted lots of interest. After
\cite{Maldacena:1998im}, there are many attempts to address the
heavy quark potential from the holography. For example, the
potential at finite temperature has been studied in \cite{AB,SJ}.
The sub-leading order correction to this quantity is discussed in
\cite{SX,ZQ}. The potential has also been investigated in some
AdS/QCD models \cite{OA2,SH1}. Other important results can be
found, for example, in \cite{JG,FB,LM,YK,SH}.

Although the theories with hyperscaling violation are intrisically
non-relativistic, we can use them as toy models for quarks from
the holography point of view. In addition, one can expect the
results that obtained from these theories shed qualitative
insights into analogous questions in QCD. In this paper, we will
investigate the heavy quark potential in the Lifshitz backgrounds
with hyperscaling violation. We want to know what will happen to
the potential if we have the quark anti-quark pair in such
backgrounds? More specifically, we would like to see how the
potential changes in the presence of the non-relativistic
parameters. In addition, we will add a constant electric field to
the backgrounds and study how it affects the potential. These are
the main motivations of the present work.

We organize the paper as follows. In the next section, the
backgrounds of the hyperscaling violation theories in \cite{EK}
are briefly reviewed. In section 3, we study the heavy quark
potential in these backgrounds in terms of the $z$ and $\theta$
parameters. In section 4, we investigate a constant electric field
effect on the heavy quark potential. The last part is devoted to
conclusion and discussion.

\section{hyperscaling violation theories}

Let us begin with a brief review of the background in \cite{EK}.
It has been argued that the Lorentz invariance is broken in this
background metric. Although charge densities induce a trivial
(gapped) behavior at low energy/temperature, there still exist
special cases where there are non-trivial IR fixed points (quantum
critical points) where the theory is scale invariant. Usually, the
metric is expressed as \cite{BG}
\begin{equation}
ds^2=u^\theta[-\frac{dt^2}{u^{2z}}+\frac{b_0du^2+dx^idx^i}{u^2}],
\end{equation}
where $b_0=\ell^2$ with $\ell$ the IR scale. The above metric is
covariant under a generalized Lifshitz scaling symmetry, that is
\begin{equation}
t\rightarrow\lambda^zt,\qquad u\rightarrow\lambda u,\qquad
x^i\rightarrow\lambda x^i,\qquad
ds^2\rightarrow\lambda^{-\theta}ds^2,
\end{equation}
where z is called the dynamical Lifshitz parameter or the
dynamical critical exponent which characterizes the behavior of
system near the phase transition. $\theta$ stands for the
hyperscaling violation exponent which is responsible for the
non-stand scaling of physical quantities as well as controls the
transformation of the metric. The scalar curvature of these
geometries is
\begin{equation}
R=-\frac{3\theta^2-4(z+3)\theta+2(z^2+3z+6)}{b_0}u^{-\theta}.
\end{equation}

The geometries are flat when $\theta=2$, $z=0,1$. The geometry is
Ricci flat when $\theta=4$, $z=3$. The geometry is in Ridler
coordinates when $\theta=0$ and $z=1$. Usually, the above special
solutions violate the Gubser bound conditions \cite{EK}. In
addition, the pure Lifshitz case is related to $\theta=0$.

By using a radical redefinition
\begin{equation}
u=(2-z)r^{\frac{1}{2-z}},
\end{equation}
and rescaling t and $x^i$, we have the following metric
\begin{equation}
ds^2\sim
r^{\frac{\theta-2}{2-z}}[-f(r)dt^2+\frac{dr^2}{f(r)}+dx^idx^i],
\qquad f(r)=f_0r^{\frac{2(1-z)}{2-z}}\label{ds},
\end{equation}
with $f_0=(2-z)^{2(1-z)}$.

In the presence of hyperscaling violations, the energy scale is
\begin{equation}
E\simeq u^{\frac{\theta-2z}{2}}\simeq
r^{\frac{\theta-2z}{2(2-z)}}.
\end{equation}

For the generalized scaling solutions of Eq.(\ref{ds}), the Gubser
bound conditions are as follows:
\begin{equation}
\frac{2z+3(2-\theta)}{2(z-1)-\theta}>0,\qquad
\frac{z-1}{2(z-1)-\theta}>0,\qquad
\frac{2(z-1)+3(2-\theta)}{2(z-1)-\theta}>0.\label{gu}
\end{equation}

Also, to consider the thermodynamic stability, one needs
\begin{equation}
\frac{z}{2(z-1)-\theta}>0.\label{gu1}
\end{equation}

More discussions about other generalized Lifshitz geometries can
be found in \cite{EK}.

The generalizations of Eq.(\ref{ds}) to include finite temperature
can be written as
\begin{equation}
ds^2\sim
(\frac{r}{\ell})^{-\alpha}[-f(r)dt^2+\frac{dr^2}{f(r)}+dx^idx^i],
\qquad f(r)=f_0(\frac{r}{\ell})^{2\beta}h, \qquad
h=1-(\frac{r}{r_h})^\gamma.\label{metric}
\end{equation}
where $\alpha=\frac{\theta-2}{z-2}$, $\beta=\frac{z-1}{z-2}$,
$\gamma=\frac{2z+3(2-\theta)}{2(2-z)}$.

The Hawking temperature is
\begin{equation}
T=\frac{f_0}{8\pi\ell}(\frac{r_h}{\ell})^{\frac{z}{z-2}}|\frac{2z-3\theta+6}{z-2}|.\label{T}
\end{equation}

\section{heavy quark potential}
In the holographic description, the heavy quark potential is given
by the expectation value of the static Wilson loop
\begin{equation}
W(C)=\frac{1}{N}Tr Pe^{i\int A_\mu dx^\mu},
\end{equation}
where C is a closed loop in a 4-dimensional space time and the
trace is over the fundamental representation of the SU(N) group.
$A_\mu$ is the gauge potential and $P$ enforces the path ordering
along the loop $C$. The heavy quark potential can be extracted
from the expectation value of this rectangular Wilson loop in the
limit $\mathcal {T}\rightarrow\infty$,
\begin{equation}
<W(C)>\sim e^{-\mathcal {T}V}.\label{w}
\end{equation}

On the other hand, the expectation value of Wilson loop in
(\ref{w}) is given by
\begin{equation}
<W(C)>\sim e^{-S_c},
\end{equation}
where $S_c$ is the regularized action. Therefore, the heavy quark
potential can be expressed as
\begin{equation}
V=\frac{S_c}{\mathcal {T}}\label{v}.
\end{equation}

We now analyze the heavy quark potential using the metric of
Eq.(\ref{metric}). The string action can reduce to the Nambu-Goto
action
\begin{equation}
S=-\frac{1}{2\pi\alpha^\prime}\int d\tau d\sigma\sqrt{-det
g_{\alpha\beta}},\label{ng}
\end{equation}
where $g$ is the determinant of the induced metric on the string
world sheet embedded in the target space, i.e.
\begin{equation}
g_{\alpha\beta}=G_{\mu\nu}\frac{\partial
X^\mu}{\partial\sigma^\alpha} \frac{\partial
X^\nu}{\partial\sigma^\beta},
\end{equation}
where $X^\mu$ and $G_{\mu\nu}$ are the target space coordinates
and the metric, and $\sigma^\alpha$ with $\alpha=0,1$ parameterize
the world sheet.

Using the parametrization $X^\mu=(t,x,0,0,r)$, $\sigma=x$,
$\tau=t$ and $r=r(x)$, we extremize the open string worldsheet
attached to a static quark at $x = +L/2$ and an anti-quark at $x =
-L/2$. Then the induced metric of the fundamental string is given
by
\begin{equation}
g_{\alpha\beta}=b(r)\left(
\begin{array}{cc}
 -f(r) & 0 \\
 0 & 1+\frac{\dot{r}^2}{f(r)}
\end{array}
\right),\label{gg}
\end{equation}
with $b(r)=(\frac{r}{\ell})^{-\alpha}$, $\dot{r}=\frac{dr}{dx}$.

Plugging (\ref{gg}) into (\ref{ng}), the Euclidean version of
Nambu-Goto action in Eq.(\ref{metric}) becomes
\begin{equation}
S=\frac{\mathcal{T}}{2\pi\alpha^\prime}\int dx
\sqrt{b^2(r)[f(r)+\dot{r}^2]}.\label{nng}
\end{equation}

We now identify the Lagrangian as
\begin{equation}
\mathcal L=\sqrt{b^2(r)[f(r)+\dot{r}^2]}.
\end{equation}

Note that $\mathcal L$ does not depend on $x$ explicitly, so the
corresponding Hamiltonian will be a constant of motion, ie.,
\begin{equation}
H=\frac{\partial\mathcal L}{\partial\dot{r}}\dot{r}-\mathcal
L=Constant=C.
\end{equation}

This constant can be found at special point $r(0)=r_c$, where
$r_c^\prime=0$, as
\begin{equation}
H=-\sqrt{f(r_c)b^2(r_c)},
\end{equation}
then a differential equation is derived,
\begin{equation}
\dot{r}=\sqrt{\frac{f^2(r)b^2(r)-f(r)f(r_c)b^2(r_c)}{f(r_c)b^2(r_c)}}\label{dotr}.
\end{equation}
with
\begin{equation}
f(r_c)=f_0(\frac{r_c}{\ell})^{2\beta}h_1, \qquad
h_1=1-(\frac{r_c}{r_h})^\gamma, \qquad
b(r_c)=(\frac{r_c}{\ell})^{-\alpha}.
\end{equation}

By integrating (\ref{dotr}) the separation length $L(\theta,z)$ of
quark-antiquark pair becomes
\begin{equation}
L(\theta,z)=2\int_0^{r_c}
dr\sqrt{\frac{f(r_c)b^2(r_c)}{f^2(r)b^2(r)-f(r)f(r_c)b^2(r_c)}}.\label{x}
\end{equation}

On the other hand, plugging (\ref{dotr}) into the Nambu-Goto
action of Eq.(\ref{nng}), one finds the action of the heavy quark
pair
\begin{equation}
S=\frac{\mathcal
{T}}{\pi\alpha'}\int_0^{r_c}drb^2(r)\sqrt{\frac{f(r)}{f(r)b^2(r)-f(r_c)b^2(r_c)}}.
\label{NG1}
\end{equation}

This action is divergent, but the divergences can be avoided by
subtracting the inertial mass of two free quarks, which is given
by
\begin{equation}
S_0=\frac{\mathcal {T}}{\pi\alpha'}\int_0^{r_h}drb(r). \label{NG2}
\end{equation}

Subtracting this self-energy, the regularized action is obtained:
\begin{equation}
S_c=S-S_0 \label{NG2}.
\end{equation}

Applying (\ref{v}), we end up with the heavy quark potential with
hyperscaling violation
\begin{equation}
V(\theta,z)=\frac{1}{\pi\alpha'}\int_0^{r_c}dr[b^2(r)\sqrt{\frac{f(r)}{f(r)b^2(r)-f(r_c)b^2(r_c)}}-b(r)]-\frac{1}{\pi\alpha'}\int_{r_c}^{r_h}drb(r),
\label{V0}
\end{equation}

Note that the potential V(z) in the Lifshitz spacetime
\cite{UH,JK} can be derived from (\ref{V0}) if we neglect the
effect of the hyperscaling violation exponent by plugging
$\theta=0$ in (\ref{V0}). Also, in the limit ($\theta=0,z=1$), the
Eq.(\ref{V0}) can reduce to the finite temperature case in
\cite{AB,SJ}.

\begin{figure}
\centering
\includegraphics[width=8cm]{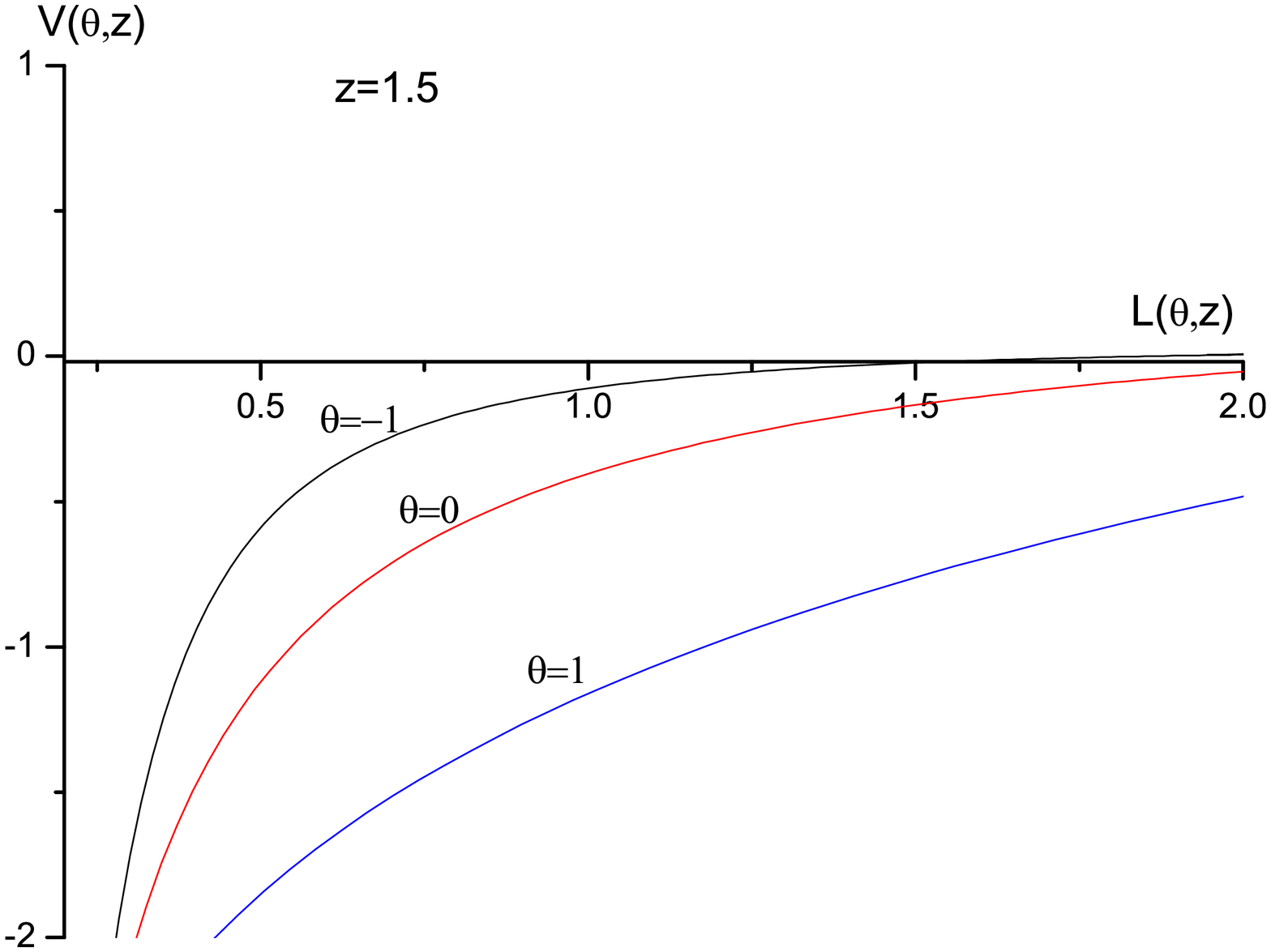}
\includegraphics[width=8cm]{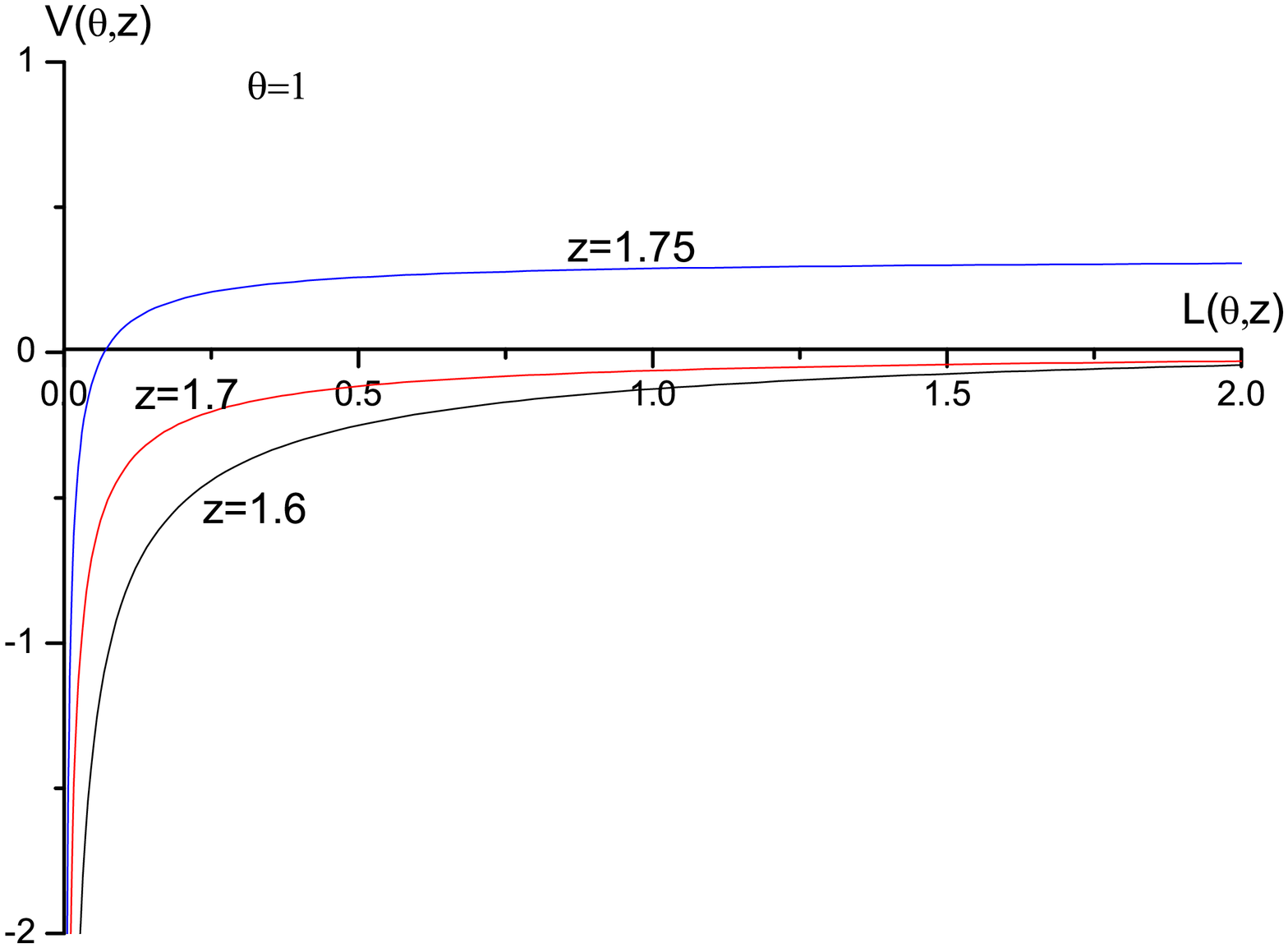}
\caption{Plots of $V(\theta,z)$ versus $L(\theta,z)$. Left:
$z=1.5$, from top to bottom $\theta=-1,0,1$ respectively; Right:
$\theta=1$, from top to bottom $z=1.75,1.7,1.6$ respectively.}
\end{figure}

Before evaluating the heavy quark potential of Eq.(\ref{V0}), we
should pause here to determine the allowed region for z and
$\theta$ at hand. The space boundary is considered at $r=0$,
consequently $\alpha>0$. To avoid the Eq.(\ref{V0}) being
ill-defined, it is required that $\gamma>0$. Moreover, one should
consider the Gubser conditions of (\ref{gu}) and the thermodynamic
stability condition of (\ref{gu1}). With these restrictions, one
finds
\begin{equation}
1<z<2, \qquad \theta<2, \qquad \theta<z,
\end{equation}
then one can choose the values of z and $\theta$ in such a range.

In Fig.1, we plot the potential $V(\theta,z)$ versus distance
$L(\theta,z)$ with different z, $\theta$. In the left plots, the
dynamical exponent is $z=1.5$ and from top to bottom the
hyperscaling violation exponent is $\theta=-1,0,1$, respectively.
In the right plots, $\theta=1$ and from top to bottom
$z=1.75,1.7,1.6$, respectively. From the figures, we can see
clearly that by increasing $\theta$ the potential decreases. One
finds also that increasing $z$ leads to increasing the potential.
In other words, increasing z and $\theta$ have different effects
on the potential. Then one can change the potential by changing
the values of these parameters. Therefore, the heavy quark
potential depends on the non-relativistic parameters.

To study how the potential changes with the temperature T, we show
$V(\theta,z)$ as a function of $L(\theta,z)$ with $z=1.6,\theta=1$
in Fig.2. From Eq.(\ref{T}), we can see that T is a decreasing
function of $z_h$ for $z=1.6$. So one finds in Fig.2 that
increasing T (or decreasing $z_h$) leads to increasing the
potential. This result is consistently with the finding of
\cite{AB,SJ}.

\begin{figure}
\centering
\includegraphics[width=10cm]{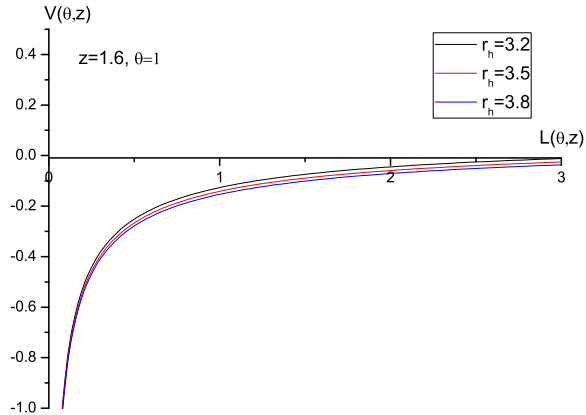}
\caption{Plots of $V(\theta,z)$ versus $L(\theta,z)$ with
$z=1.6,\theta=1$. From top to bottom $r_h=3.2,3.5,3.8$
respectively.}
\end{figure}

Moreover, to see the short distance behavior of the potential, we
take the limit $L(\theta,z)T\rightarrow0$ and find the following
approximate formula
\begin{equation}
L(\theta,z)\simeq
\frac{2f_0^{-1/2}}{\alpha-2\beta+1}r_c^{1-\beta},\qquad
V(\theta,z)\simeq
\frac{\sqrt{\lambda}}{\pi(1-\alpha)}r_c^{1-\alpha},
\end{equation}
which yields
\begin{equation}
V(\theta,z)\propto L(\theta,z)^{\theta-z}.
\end{equation}

One can see that the potential is dependent on $z$ and $\theta$.
In the special case of $\theta=0,z=1$ (or $\alpha=2,\beta=0$), one
finds that the potential is of Coulomb type:
\begin{equation}
V(\theta,z)\simeq -\frac{\sqrt{\lambda}}{1.5\pi L(\theta,z)},
\end{equation}
but for other cases, the potential may not be Coulombian.

\section{Effect of constant electric field}

In this section, we study the effect of a constant electric field
on the heavy quark potential following the method proposed in
\cite{TM}. The constant B-field is along the $x^1$ and $x^2$
directions. As the field strength is involved in the equations of
motion, this ansatz could be a good solution to supergravity as
well as a simple way of studying the B-field correction. The
constant B-field is added to the metric of Eq.(\ref{metric}) by
the following form:
\begin{equation}
B=B_{01}dt\wedge dx^1+B_{12}dx^1\wedge dx^2,
\end{equation}
where $B_{01}$ and $B_{12}$ are assumed to be constants with
$B_{01}=E$ the NS-NS antisymmetric electric field and $B_{12}=H$
the NS-NS antisymmetric magnetic field.

The constant B-field considered here is only turned on $x^1$
direction, which implies $H=0$. After adding an electric field to
this background metric of Eq.(\ref{metric}), the string action is
given by
\begin{equation}
S=-\frac{1}{2\pi\alpha^\prime}\int d\tau d\sigma\sqrt{-det
(g+b)_{\alpha\beta}},\label{ng1}
\end{equation}
where $g_{\alpha\beta}$ is given in Eq.(\ref{gg}).
$b_{\alpha\beta}=B_{\mu\nu}\partial_\alpha X^\mu\partial_\beta
X^\nu$, is obtained as
\begin{equation}
b_{\alpha\beta}=\left(
\begin{array}{cc}
 0 & 0\\
 0 & \xi
\end{array}
\right),\label{gg1}
\end{equation}
then the string action in Eq.(\ref{ng1}) reads
\begin{equation}
S=\frac{\mathcal{T}}{2\pi\alpha^\prime}\int dx
\sqrt{b^2(r)f(r)+b(r)f(r)\xi+b^2(r)\dot{r}^2}.\label{nng1}
\end{equation}

Parallel to the case of the previous section, we have
\begin{equation}
\dot{r}=\sqrt{\frac{A^2(r)-A(r)A(r_c)}{A(r_c)b^2(r)}}.
\end{equation}
with
\begin{equation}
A(r)=b^2(r)f(r)+b(r)f(r)\xi, \qquad
A(r_c)=b^2(r_c)f(r_c)+b(r_c)f(r_c)\xi.
\end{equation}

We call again the separation length and the heavy quark potential
as $L$ and $V$, respectively. One finds
\begin{equation}
L=2\int_{0}^{r_c}dr\sqrt{\frac{A(r_c)b^2(r)}{A^2(r)-A(r)A(r_c)}},\label{x1}
\end{equation}
and
\begin{equation}
V=\frac{1}{\pi\alpha'}\int_{0}^{r_c}dr[\sqrt{\frac{A(r)b^2(r)}{A(r)-A(r_c)}}-b(r)]-\frac{1}{\pi\alpha'}\int_{r_c}^{r_h}drb(r).
\label{V1}
\end{equation}

\begin{figure}
\centering
\includegraphics[width=10cm]{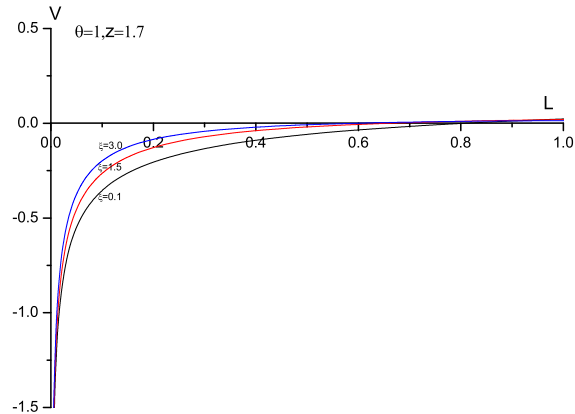}
\caption{Plots of $V$ versus $L$ with different $\xi$. Here
$z=1.7$, $\theta=1$. From top to bottom $\xi=3.0,1.5,0.1$.}
\end{figure}

To see the effect of the constant electric field $\xi$ on the
heavy quark potential in the backgrounds with hyperscaling
violation. We plot the heavy quark potential as a function of the
inter distance for $z=1.7$, $\theta=1$ with three different $\xi$
in Fig.3. In the plots from top to bottom $\xi=3.0,1.5,0.1$
respectively. From the figures, we can see that the heavy quark
potential increases with increasing $\xi$. In other words, the
presence of the constant electric field leads to a smaller
screening radius. This result can be understood by considering the
relation between the potential and the viscosity of the medium. It
was argued \cite{JN} that increasing the viscosity, the screening
of the potential due to the thermal medium weakens and so the
potential decreases. On the other hand, the presence of the
constant electric field tends to weaken the viscosity \cite{TM}.
Thus, increasing the constant electric field leads to weakening
the viscosity or increasing the potential.

\section{conclusion and discussion}

In this paper, we have investigated the heavy quark potential in
the backgrounds with hyperscaling violation at finite temperature.
These theories are strongly coupled with anisotropic scaling
symmetry in the time and a spatial direction. Although the
theories are not directly applicable to QCD, the features of them
are similar to QCD. Therefore one can expect the results that
obtained from these theories shed qualitative insights into
analogous questions in QCD. In addition, an understanding of how
the heavy quark potential changes by these theories may be useful
for theoretical predictions.

In section 3, we used the holographic description to calculate the
heavy quark potential at finite temperature. We considered the
space boundary at $r=0$ and discussed the potential for a certain
range of $z$ and $\theta$ which satisfies the Gubser conditions
and the thermodynamic stability condition. In is shown that
increasing z and $\theta$ have different effects: the potential
increases as z increases but it decreases as $\theta$ increases.
As a result, the heavy quark potential depends on the
non-relativistic parameters. In section 4, we added a constant
electric field to the background metrics and study its effect on
the heavy quark potential. We observed that the potential rises as
the constant electric field increases. In other words, the
presence of the constant electric field leads to increasing the
heavy quark potential.

Finally, it is interesting to mention that the drag force
\cite{CP} can also be studied in the backgrounds with hyperscaling
violation. We will leave this for further study.

\section{Acknowledgments}

This research is partly supported by the Ministry of Science and
Technology of China (MSTC) under the ¡°973¡± Project no.
2015CB856904(4). Zi-qiang Zhang and Gang Chen are supported by the
NSFC under Grant no. 11475149. De-fu Hou is supported by the NSFC
under Grant no. 11375070, 11521064.


\end{document}